\ifx\documentclass\undefined
\documentstyle[12pt]{article}
\else
\documentclass[12pt]{article}
\fi

\makeatletter
\renewcommand{\theequation}{\thesection.\arabic{equation}}
\@addtoreset{equation}{section}
\def\eqnarray{%
\stepcounter{equation}%
\let\@currentlabel=\theequation
\global\@eqnswtrue
\global\@eqcnt\z@
\tabskip\@centering
\let\\=\@eqncr
$$\halign to \displaywidth\bgroup\@eqnsel\hskip\@centering
$\displaystyle\tabskip\z@{##}$&\global\@eqcnt\@ne
\hfil$\displaystyle{{}##{}}$\hfil
&\global\@eqcnt\tw@$\displaystyle\tabskip\z@{##}$\hfil
\tabskip\@centering&\llap{##}\tabskip\z@\cr}
\makeatother

\def\bbbz{{\mathchoice {\hbox{$\sf\textstyle Z\kern-0.4em Z$}}
{\hbox{$\sf\textstyle Z\kern-0.4em Z$}}
{\hbox{$\sf\scriptstyle Z\kern-0.3em Z$}}
{\hbox{$\sf\scriptscriptstyle Z\kern-0.2em Z$}}}}
\def\bbbq{{\mathchoice {\setbox0=\hbox{$\displaystyle\rm Q$}\hbox{\raise
0.15\ht0\hbox to0pt{\kern0.4\wd0\vrule height0.8\ht0\hss}\box0}}
{\setbox0=\hbox{$\textstyle\rm Q$}\hbox{\raise
0.15\ht0\hbox to0pt{\kern0.4\wd0\vrule height0.8\ht0\hss}\box0}}
{\setbox0=\hbox{$\scriptstyle\rm Q$}\hbox{\raise
0.15\ht0\hbox to0pt{\kern0.4\wd0\vrule height0.7\ht0\hss}\box0}}
{\setbox0=\hbox{$\scriptscriptstyle\rm Q$}\hbox{\raise
0.15\ht0\hbox to0pt{\kern0.4\wd0\vrule height0.7\ht0\hss}\box0}}}}
\def\bbbc{{\mathchoice {\setbox0=\hbox{$\displaystyle \rm C$}\hbox{\raise
0.06\ht0\hbox to0pt{\kern0.4\wd0\vrule height0.9\ht0\hss}\box0}}
{\setbox0=\hbox{$\textstyle\rm C$}\hbox{\raise
0.06\ht0\hbox to0pt{\kern0.4\wd0\vrule height0.9\ht0\hss}\box0}}
{\setbox0=\hbox{$\scriptstyle\rm C$}\hbox{\raise
0.06\ht0\hbox to0pt{\kern0.4\wd0\vrule height0.8\ht0\hss}\box0}}
{\setbox0=\hbox{$\scriptscriptstyle\rm C$}\hbox{\raise
0.06\ht0\hbox to0pt{\kern0.4\wd0\vrule height0.8\ht0\hss}\box0}}}}
\makeatletter
  \renewcommand{\theequation}{%
 \thesection.\arabic{equation}}
  \@addtoreset{equation}{section}
\makeatother
\newtheorem{theorem}{Theorem}[section]
\newtheorem{lemma}{Lemma}[section]

\newtheorem{remark}{Remark}[section]

\newsavebox{\toy}
\savebox{\toy}{\framebox[0.65em]{\rule{0cm}{1ex}}}
\newcommand{\QED}{\usebox{\toy}}
\def\nlni{\par\ifvmode\removelastskip\fi\vskip\baselineskip\noindent}

\begin{document}
\setlength{\baselineskip}{15pt}
\title{
Spin of the ground state
and 
the flux phase problem on the ring
}
\author{Fumihiko Nakano\thanks{
Mathematical Institute, Tohoku University, 
Sendai, 980-8578, Japan. 
This work is partially supported by JSPS grant 
15740049.}}
\date{}
\maketitle
\begin{abstract}
As a 
continuation of our previous work, 
we derive the optimal flux phase which minimizes the ground state energy in the one-dimensional  many particle systems, when the number of particles is odd in the absence of on-site interaction and external potential.  
Moreover, 
we study the relationship between the flux on the ring and the spin of the ground state through which we derive some information on the sum of the lowest eigenvalues of one-particle Hamiltonians. 
\end{abstract}

Short tittle: Spin and flux on the ring \\


\section{Introduction}
The flux phase problem 
is to derive the optimal flux distribution which minimizes the ground state energy of the system of many fermions. 
There are a few physical 
significances of this problem, and one of which  is that the diamagnetic inequality, which widely holds for one-particle Hamiltonians, is sometimes reversed for many particle ones. 
As for the mathematical results, 
we refer to \cite{LL, LN, L2} 
where many cases are studied at half-filling for bipartite rings, lattices, and ones with some particular geometry such as tree of rings and hidden trees. 
Bethe-ansatz calculations
are done in \cite{YF} where they study whether the current response to the variation of the magnetic flux is diamagnetic or paramagnetic. 
In this paper, 
we continue our study to derive the optimal flux of the Hubbard Hamiltonian on the ring 
$\Lambda :=
\{ 1, 2, \cdots, L \}$ ($L+1 \equiv 1$) 
defined by 
\[
H :=
\sum_{\sigma = \uparrow, \downarrow}
\sum_{x=1}^L 
t_{x, x+1} c_{x+1, \sigma}^{\dagger} c_{x, \sigma}
+(h.c.)
+
\sum_{\sigma = \uparrow, \downarrow}
\sum_{x=1}^L 
V(x) n_{x, \sigma}
+
\sum_{x=1}^L U(x) n_{x, \uparrow} n_{x, \downarrow}
\]
where 
$c_{x,\sigma} (c_{x, \sigma}^{\dagger})$
are the annihilation (creation) operator satisfying the canonical anticommutation relations and 
$n_{x, \sigma} := c_{x, \sigma}^{\dagger} c_{x, \sigma}$. 
$t_{x, x+1} \ne 0$
and 
$\arg t_{x, x+1} = \theta_x \in [0, 2\pi)$ 
such that 
$\sum_{x=1}^L \theta_x = \varphi$ 
(mod $2\pi$).
$U(x), V(x) \in {\bf R}$. 
Eigenvalues of 
$H$
is independent of the choice of 
$\{ \theta_x \}_{x=1}^L$ 
such that 
$\sum_{x=1}^L \theta_x = \varphi$ 
so that we write 
$H = H(\varphi)$. 
We consider 
$H(\varphi)$ 
on the spin 
$\frac 12$ 
$N$-fermion Hilbert space 
${\cal H}_N$ 
which is the span of 
\[
B_N
:=
\left\{
c_{x_1, \sigma_1}^{\dagger}
c_{x_2, \sigma_2}^{\dagger}
\cdots
c_{x_N, \sigma_N}^{\dagger}
| \mbox{vac}> : 
x_j \in \Lambda, \sigma_j = \uparrow, \downarrow, 
j=1, 2, \cdots, N 
\right\}.
\]
Let 
$E_N (\varphi)$ 
be the ground state energy of 
$H(\varphi)$ : 
\[
E_N (\varphi) 
:= \min 
\left\{ < \Phi, H(\varphi) \Phi> : 
\Phi \in {\cal H}_N, < \Phi, \Phi > = 1 
\right\}.
\]
Our aim 
is to derive the optimal flux 
$\varphi_{opt}$ 
which minimizes 
$E_N (\varphi)$ : 
$E_N (\varphi_{opt}) 
=
\min_{\varphi \in [0, 2 \pi)} E_N (\varphi)$. 
Uniqueness of 
$\varphi_{opt}$, 
which is not discussed in this paper, 
holds when 
$T := \{ | t_{x, x+1} | \}_{x=1}^L$ 
has some periodicity, or 
$T$ 
and 
$V$ 
satisfy some particular relation \cite{NN}. 
In \cite{N}, 
we studied the case where 
$N$ 
is even.
The result there was : 
\begin{theorem}
{\bf (Optimal flux on the ring: even case)}\\
Let $N \le L$
be even. \\
(1) $U < \infty$: 
$\varphi_{opt}=\left(\frac N2+1\right)\pi$
($L$ is even)
$=\frac {N\pi}{2}$
($L$ is odd).\\
(2) $U = \infty$: 
$\varphi_{opt} = \frac {2n}{N}\pi$,
$n=0,1, \cdots, N-1$.
\label{even}
\end{theorem}
The key ingredient 
of the proof of Theorem \ref{even} was to regard 
$H(\varphi)$
as a hopping Hamiltonian on 
$B_N$
and compute the flux through the circuit in 
$B_N$ 
of `minimal' length. 
The distinction 
between 
$0$
and 
$\pi$ 
comes from counting how many times a particle exchanges its location with others in these circuits.
When 
$U=\infty$, 
such exchanges are not possible and hence there is no distinction. 
In fact,
$E_N^{\infty} (\varphi):=\lim_{U \uparrow \infty}
E_N(\varphi)$
has period
\footnote{This fact and its implications are discussed by 
\cite{K,YF}.}
$\frac {2 \pi}{N}$
and 
$H_{\infty}(0)$ 
is gauge equivalent to  
$H_{\infty}(\pi)$
\footnote{
$H_{\infty} (\varphi) := P H(\varphi) P$
and 
$P := \prod_{x \in \Lambda} ( 1 - n_{x, \uparrow}n_{x, \downarrow})$
is the orthogonal projection onto the space of states with no doubly occupied sites. }.

We turn to the case where 
$N$ 
is odd and 
$U=0$. 
Some computations 
of examples imply that 
$\varphi_{opt}$ 
depends on 
$U$
in general and there seems to be no general rule except the half-filling case. 
\begin{theorem}
{\bf (Optimal flux on the ring: odd case)}\\
Let 
$N = L$ 
be odd and 
$U=V=0$. 
Then 
$E_N (\varphi)$ 
has period 
$\pi$ 
and is minimized if 
$\varphi = \frac {\pi}{2}, \frac {3\pi}{2}$. 
\label{odd}
\end{theorem}
\begin{remark}
The same result is deduced in \cite{R}
by a different argument.
For the 
translation invariant case
($t_{x,x+1}, U_x$
are constant), 
Bethe-ansatz calculation has been done 
\cite{YF}
and the result in Theorem \ref{odd} is the same as 
they obtained. 
Since 
we set $U=0$, 
only the free particle case is considered in Theorem \ref{odd}. 
So our contribution is that 
the hopping coefficients 
$T = \{ t_{x,x+1} \}_{x=1}^L$
can be arbitrary which is not covered by the Bethe-ansatz solutions. 
Therefore, in free case, 
the hopping disorder has no effect on the optimal flux. 
\end{remark}
\begin{remark}
If 
$U=\infty$
and 
$N (< L)$
is odd, 
the argument of the proof of Theorem \ref{even}(2)
proves that 
$E(\varphi)$
has period 
$\frac {2 \pi}{N}$
and 
$\varphi_{opt} = 
\frac {2n}{N} \pi$ 
($L$ even), 
$\frac {2n+1}{N} \pi$
($L$ odd), 
$n=0, 1, \cdots, N-1$. 
\end{remark}
\begin{remark}
The following example implies the conclusion of 
Theorem \ref{odd}
is not true in general if 
$V \ne 0$
so that the potential disorder may have some effect on the optimal flux.
Let 
$N=L=5$
and let
\[
| t_{x, x+1} |=\cases{
1, & $(x =1,4)$ \cr
t, & $(x =3)$ \cr
\sqrt{2}, & $(x =2,4),$ \cr}
\quad
V(x) = \cases{
0, & $(x \ne 3,4)$\cr
t, & $(x =3,4)$\cr}, 
\quad
t > 0. 
\]
Since 
the Hamiltonian 
$H(\varphi)$
contains terms of the form 
$t (c_{3, \sigma}^{\dagger}+c_{4, \sigma}^{\dagger})
(c_{3, \sigma}+c_{4, \sigma})$,
when 
$t$
is sufficiently large, eigenvalues of 
$H(\varphi)$ 
approach to that of 
$H'(\varphi + \pi)$ 
in which 
$N=5, L=4$
and 
$| t_{x,x+1} | = 1$
for any 
$x$. 
The ground state energy of 
$H'(\varphi + \pi)$ 
is minimized if and only if
$\varphi = \pi \pm 4 \arcsin \frac {1}{\sqrt{5}}$. 
On the other hand, 
we believe 
Theorem \ref{odd}
is true when
$U\ne 0$ 
as the computations in translation invariant cases imply
\cite{YF}.
\label{potential disorder}
\end{remark}
\begin{remark}
At finite temperature,
optimal flux is different from 
$\frac {\pi}{2},\frac {3\pi}{2}$
in general. 
In fact, in the canonical ensemble, 
the partition function
$P(\varphi):=\mbox{Tr }[e^{-\beta H(\varphi)}]$
(restricted on 
$S_z = \frac 12$
subspace for simplicity)
is a complicated function of 
$\varphi$
if 
$\beta$
is large, and 
$\varphi = \frac {\pi}{2}, \frac{3\pi}{2}$
does not necessarily maximize it,
although
they are always the critical points. 
This is different from 
the case of even number of particles where 
$P(\varphi)$
is maximized for any 
$\beta > 0$ 
by the optimal flux given in Theorem \ref{even}
\cite{N}.
In the grand canonical ensemble, 
the average particle number depends on 
$\varphi, \beta$
and the absolute ground state does not lie at half-filling unless
$\varphi = \frac {\pi}{2}, \frac {3 \pi}{2}$. 
In \cite{LL}, 
it is shown that the grand canonical partition function with zero chemical potential is maximized if 
$\varphi = 0, \pi$.
\label{nonzero temperature}
\end{remark}
Next, 
we study the spin of the ground state.
In what follow, 
we assume 
$L$
is even for simplicity;
the results for odd
$L$ 
follow by exchanging 
$0$
and
$\pi$
in each statement of theorems given below.
The proof of 
Theorem \ref{even}, together with the Lieb-Mattis argument \cite{LM} 
proves the following fact\footnote{Theorem \ref{singlet}
is pointed out by professor E. Lieb to whom the author is grateful.}.
\begin{theorem}
{\bf (Ground state is unique with spin zero)}\\
Let 
$U < \infty$, 
$N$ 
even and 
$\varphi = \left(\frac N2+1 \right) \pi$ 
(mod $2\pi$).
Then 
the ground state of 
$H(\varphi)$
is unique and 
$S=0$. 
\label{singlet}
\end{theorem}
\begin{remark}
If 
$\varphi = \frac {N\pi}{2}$ (mod 2$\pi$)
and 
$|t_{x,x+1}| = 1, U=V=0$, 
then the ground state of 
$H(\varphi)$ 
is not unique and 
$S=0, 1$. 
This contrasts with 
Lieb-Mattis theorem 
\cite{LM} 
which states that the ground state is always unique and 
$S=0$ 
in the one-dimensional chain with open boundary condition
(and thus no flux is present so that one can freely adjust the sign of the matrix elements). 
The example above 
shows, if 
$\varphi$
is not optimal, the boundary effect is not negligible in general. 
We also remark that such 
`non-unique' situation is not stable under the variation of 
$T,V$, and $U$. 
For instance, once 
$U_x < 0$ 
for any 
$x$, 
then the ground state is again unique and 
$S=0$
\cite{L1}. 
On the other hand, 
Theorem \ref{singlet}
says, if 
$\varphi$
is optimal, this uniqueness property is stable which holds for any 
$T,V$ and $U$.
\end{remark}
\begin{remark}
When 
$N=L$
is odd, 
$U=V=0$,
 and 
$\varphi= \frac {\pi}{2}, \frac {3\pi}{2}$, 
then the ground state is unique with 
$S=\frac 12$
apart from the 
$(2S+1)$-degeneracy.
\label{odd spin}
\end{remark}
When 
$U=\infty$, 
there are some relationship between the flux 
$\varphi$
and the spin of the ground state.
Let 
$\{ e_j (\varphi) \}_{j=1}^L$ 
be the eigenvalue (in increasing order) of the one-particle Hamiltonian 
$h(\varphi)$
corresponding to 
$H(\varphi)$
(that is, 
$H(\varphi)$ 
as an operator on
${\cal H}_1$).
\begin{theorem}
{\bf (Spin and flux are related)}\\
Let 
$N(< L)$
be even and 
$U=\infty$. \\
(1)
$H_{\infty} (0)$
does not have the ground state with 
$S=\frac N2$
if and only if 
$\sum_{j=1}^N e_j (\pi) < \sum_{j=1}^N e_j (0)$. \\
(2)
$H_{\infty} (0)$ 
does not have the ground state with 
$S=\frac N2$.\\
\label{relation}
\end{theorem}
\begin{remark}
Theorem \ref{relation}
implies that the spin of the ground state changes when the flux changes.
For instance, let 
$N=4n+2$. 
Then
$H_{\infty} (\pi)$ 
has a ground state with 
$S=\frac N2$
while 
$H_{\infty} (0)$
does not, but have one with 
$S=0$. 
\end{remark}
\begin{remark}
The inequality 
$\sum_{j=1}^N e_j (\pi) \le \sum_{j=1}^N e_j (0)$
follows from Theorem \ref{even}.
So the statement
$\sum_{j=1}^N e_j (\pi) < \sum_{j=1}^N e_j (0)$
has something to do with the uniqueness question of the optimal flux. 
Theorem \ref{relation}
says that an ``analytical" statement 
$\sum_{j=1}^N e_j (\pi) < \sum_{j=1}^N e_j (0)$
is equivalent to a property of the spin of the ground state, which is robust under the variation of 
$T,V$, and $U$. 
\end{remark}
Finally, 
we discuss an connection between the ferromagnetic
($S=\frac N2$)
ground state of 
$H_{\infty}(\pi)$
and the singlet
($S=0$) one of 
$H_{\infty}(0)$. 
Since 
$H_{\infty}(\pi)$
is gauge equivalent to 
$H_{\infty}(0)$, 
there is a gauge transformation 
$g$ 
under which 
$H_{\infty}(\pi)$
is transformed to 
$H_{\infty}(0)$
\footnote{$g$
is not unique, since 
$H_{\infty}(\varphi)$
is not irreducible.}. 
Because 
the ground state of 
$H_{\infty}(\pi)$
is degenerate
(it has at least all even(odd) spins for 
$N=4n$($4n+2$)), 
it is not clear how each ground state of 
$H_{\infty}(\pi)$
is transformed under 
$g$. 
In fact, when 
$N = 4n$, 
the ground states of 
$H_{\infty}(0)$
can have all spins such that 
$S< \frac N2$
and 
$g \Psi_f^{\pi, \infty}$
does not have fixed spin. 
However, if 
$N=4n+2$, 
we have the following theorem, which says that the 
ferromagnetic ground state of 
$H_{\infty}(\pi)$
is directly connected to the singlet ground state of 
$H_{\infty}(0)$ 
via the gauge transformation mentioned above. 
\begin{theorem}
{\bf (A connection between ferromagnetic and singlet states)}\\
Let 
$N=4n+2$
and let 
$\Psi_f^{\pi, \infty}$
be the ferromagnetic ground state of 
$H_{\infty}(\pi)$.
Then there is a gauge transformation 
$g_{\infty}$
under which 
$H_{\infty}(\pi)$
is transformed to
$H_{\infty}(0)$
and 
$g_{\infty} \Psi_f^{\pi, \infty}$
is a singlet ground state of 
$H_{\infty}(0)$. 
\label{connection}
\end{theorem}
The singlet state 
$g_{\infty}\Psi_f^{\pi, \infty}$
is described as follows. 
If we write 
$\Psi_f^{\pi, \infty}$
as a linear combination of elements of 
$B_N$, 
coefficients are the same for every configurations of spins for each fixed locations of particles. 
The gauge transformation 
$g_{\infty}$
then puts 
$(-1)$
alternately on every cyclic permutation of spins. 
Therefore, 
the singlet ground state of 
$H_{\infty} (0)$
is a sort of `spiral' state in the configuration space 
$B_N$ 
produced from the ferromagnetic one. 

In section 2, 
we give proof of theorems.
Theorem \ref{odd}
is proved by reducing the problem to the case of even number of particles using the ideas of Floquet analysis.
We remark that 
a simple adaptation of the method of proof of 
Theorem \ref{even} would lead us to a complicated computation of the partition function
$P(\varphi)$
of 
$H(\varphi)$. 
Theorem \ref{singlet} 
is proved by putting the arguments in \cite{LM, N} together. 
The key fact is that 
the ground state of $H_{U\ne 0}$ and $H_{U=0}$ are both unique and not orthogonal to each other. 
The ground state of 
$H_{U=0}$
has spin zero because it is unique. 
To prove Theorem \ref{relation}(1), 
we use 
Perron-Frobenius theorem which implies that 
$H_{\infty} (\pi)$ 
has the ferromagnetic state which makes it possible to derive the ground state energy of 
$E_N (\pi)$, 
which is equal to 
$E_N(0)$ 
since 
$H_{\infty} (0)$ 
and 
$H_{\infty} (\pi)$
are gauge equivalent. 
Then 
the equivalence follows from comparing ferromagnetic energies of 
$H_{\infty} (0)$
and  
$H_{\infty} (\pi)$.
Theorem \ref{relation}(2)
follows from comparing the spin of the ground state of 
$H_{\infty} (0)$ 
with that of 
$H_{\infty}^0(0)$
where 
$|t_{x,x+1}|=1$
and 
$V=0$. 
To prove Theorem \ref{connection}, 
we note that for 
$U<\infty$, $H(0)$
is gauge equivalent to 
$H_{PF}$ 
whose matrix elements 
($B_N$ 
as its basis)
are non-positive.
Ground states of both
are unique and that of 
$H(0)$
has 
$S=0$ 
while one of 
$H_{PF}$
is positive\footnote{A state 
$\Psi$
is positive(non-negative) means that 
$\Psi$
is expanded as
$\Psi = \sum_j a_j \psi_j$, 
$\psi_j \in B_N$
with 
$a_j > 0$($a_j \ge 0$)
for all 
$j$. }.
When 
$U$
goes to infinity, the ground state of 
$H(0)$
tends to the singlet one of 
$H_{\infty}(0)$
while the ground state of 
$H_{PF}$
tends to the ferromagnetic one of 
$H_{\infty}(\pi)$. 

Section 3 
is devoted to the discussion, and 
in Appendix, we prove a simple lemma which 
appears in the proof of Theorem \ref{even}(2).
%

%
\section{Proof of Theorems}
First of all, 
we provide the proof of Theorem \ref{even}(2)
for the sake of completeness, 
because in \cite{N}, we only asserted 
$\varphi_{opt}=0, \pi$. \\

{\it Proof of Theorem \ref{even}(2)}
We assume 
$L$ 
is even; the proof for odd
$L$
follows similarly. 
We always work on 
$S_z=0$
subspace of 
${\cal H}_N$
and let 
${\cal G} =\mbox{Range }P$
be the space of states with no doubly occupied sites.
Let 
${\cal G}={\cal G}_1 \oplus {\cal G}_2 \oplus \cdots \oplus {\cal G}_K$
be the decomposition of 
${\cal G}$
such that 
$H_j(\varphi) := H(\varphi) |_{{\cal G}_j}$
is irreducible.
We choose the basis 
$B_j$ 
of 
${\cal G}_j$
as
\begin{equation}
B_j := \left\{
c_{x_1, \sigma_1}^{\dagger}
c_{x_2, \sigma_2}^{\dagger}
\cdots
c_{x_N, \sigma_N}^{\dagger}
|\mbox{vac}>
\; : \;
x_1 < x_2 < \cdots < x_N, 
\;
\sigma_j = \uparrow, \downarrow 
\right\}.
\label{b}
\end{equation}
Since 
$U = \infty$, 
exchange of particles is not allowed so that for each
$c_{x_1, \sigma_1}^{\dagger}
c_{x_2, \sigma_2}^{\dagger}
\cdots
c_{x_N, \sigma_N}^{\dagger}
|\mbox{vac}> \in B_j$, 
the spin configuration 
$( \sigma_1, \sigma_2, \cdots, \sigma_N)$ 
of that can be obtained by the cyclic permutation
\footnote{The one-times cyclic permutation of a configuration
$(\tau_1, \tau_2, \cdots, \tau_N)$
is defined by 
$(\tau_2, \tau_3, \cdots, \tau_N, \tau_1)$.}
 of a fixed spin configuration 
$(\tau_1, \tau_2, \cdots, \tau_N)$. 
There exists 
$p (=2, \cdots, N)$
such that 
$(\tau_1, \tau_2, \cdots, \tau_N)$
is  invariant under the cyclic permutations of $p$-times. 
Because 
we are working in 
$S_z = 0$
subspaces, 
$p$
must be even. 
In this case,
we say 
${\cal G}_j$
has period 
$p$.
We rearrange 
${\cal G}_j$'s
w.r.t. their period and rewrite,
${\cal G}= \oplus_{p=2}^N \oplus_{j=1}^{J_p} {\cal G}_j^p$, 
where 
${\cal G}_j^p$
has period 
$p$ 
with 
$B_j^p$
as its basis which is chosen like 
(\ref{b}). 
Let 
$H_j^p (\varphi) := H(\varphi) |_{{\cal G}_j^p}$
which we regard as a hopping Hamiltonian on 
$B_j^p$. 
The flux 
$\Phi_j^p$
of these circuits in 
$B_j^p$
with `minimal' length\footnote{`Minimal'
means circuits having least length whose flux depends on 
$\varphi$. 
If 
$L \ge 4$, 
the length of circuits of least length is always 
$4$, 
but fluxes there are always zero and do not affect the discussion here. }
is given by 
\[
\Phi_j^p
=
p\varphi + p(N-1) \pi 
\equiv 
p \varphi
\quad
(\mbox{mod }2 \pi).
\]
The first term 
comes from the hopping of particles and the second one comes from the fact that if a particle hops from the site $L$ to the site $1$, we have to add $\pi$ to the flux
(as discussed in the proof of Theorem in \cite{N}). 
Therefore 
the lowest eigenvalue 
$E_j^p (\varphi)$
of 
$H_j^p (\varphi)$
is minimized if 
$\varphi_j^p = \frac {2 \pi n}{p}$, 
$n=0, 1, \cdots, p-1$. 
Since 
$p$
is even, they always include 
$0, \pi$.
Hence 
\begin{equation}
E_j^p (\pi)
=
E_j^p \left(
\frac {2 \pi n}{p}
\right), 
\quad
n=0, 1, \cdots, p-1.
\label{one}
\end{equation}
If 
$\varphi = \pi$, 
by taking the gauge such that 
$t_{x, x+1} < 0$
($x = 1, 2, \cdots, N-1$), 
and 
$t_{N,1} > 0$, 
the matrix elements of 
$H_j^p(\varphi)$
in terms of the basis 
$B_j^p$ 
are non-positive.
Hence, 
by Perron-Frobenius theorem, 
we have a ferromagnetic ground state 
$\Psi_{f}$
of 
$H_{\infty}(\pi)$
so that for some 
$\{ a_j^p \}_{j, p}$, 
it is written as
\begin{equation}
\Psi_{f}
=
\sum_{j,p}^K a_j^p \psi_j^p
\label{f}
\end{equation}
where 
$\psi_j^p$
is the lowest eigenvector of 
$H_j^p (\pi)$. 
Since 
it has maximal spin, it can also be written as
\[
\Psi_{f}
=
\sum_{x_1, \cdots, x_N}
b_{x_1, \cdots, x_N}
\sum_{\sigma_1, \cdots, \sigma_N}
c_{x_1, \sigma_1}^{\dagger}
c_{x_2, \sigma_2}^{\dagger}
\cdots
c_{x_N, \sigma_N}^{\dagger}
| \mbox{vac }>
\]
with
$b_{x_1, \cdots, x_N} > 0$. 
Therefore 
for any fixed 
$x_1, x_2, \cdots, x_N$, 
every spin configuration 
$(\sigma_1, \sigma_2, \cdots, \sigma_N)$
appears in 
(\ref{f}), 
so that 
$a_j^p \ne 0$
for any 
$j, p$. 
Hence 
the lowest eigenvalue 
$E_j^p (\pi)$
are the same for any $j, p$: 
\begin{equation}
E_j^p (\pi) 
=
E_N^{\infty}(\pi).
\label{two}
\end{equation}
By the diamagnetic inequality, we have 
\begin{equation}
E_j^p (\pi) \le E_j^p (\varphi), 
\quad
\varphi \in [0, 2 \pi].
\label{three}
\end{equation}
The assertion 
$\varphi_{opt} = \frac {2 \pi n}{N}$
then follows from 
(\ref{one}), (\ref{two}), (\ref{three}). 

The claim that 
$E(\varphi)$
has period 
$\frac {2 \pi}{N}$ 
is proved by the following lemma.
\begin{lemma}
$E_k^N (\varphi) \le E_j^p (\varphi)$ 
for any 
$p= 2, \cdots, N$
and 
$j=1, 2, \cdots, J_p$.
\label{boson}
\end{lemma}
The proof 
of Lemma \ref{boson} is given in the appendix for completeness, 
which is a simple proof of the fact : 
`the hard core boson has the lowest energy'. 
Lemma \ref{boson} 
also gives an alternative and simpler proof of Theorem \ref{even}(2), 
for 
$E_j^N (\varphi)$
has period 
$\frac {2 \pi}{N}$. 
\QED\\

{\it Proof of Theorem \ref{odd}}
Let 
$\{ e_j (\varphi) \}_{j=1}^L$ 
be the eigenvalue (in increasing order) of
$H(\varphi)$ 
on
${\cal H}_1$,
that is, eigenvalues of the corresponding one-particle Hamiltonian 
$h(\varphi)$, 
and let 
$F_K (\varphi) := \sum_{j=1}^K e_j (\varphi)$
be the sum of the 
$K$
lowest eigenvalues.
Let 
$N=2n +1$.
By hole-particle transformation for down spins 
and 
by the assumption that 
$V=0$, 
we have
$E_N (\varphi) = F_n (\varphi) + F_{n+1}(\varphi)
=
F_n (\varphi) + F_n (\varphi + \pi)$.
In what follows we show 
\begin{equation}
F_n (\varphi) + F_n (\varphi+\pi)
=
F_{2n}^{2L}(2 \varphi)
\label{double}
\end{equation}
where 
$F_K^{2L}(\varphi)$
is the sum of the 
$K$
lowest eigenvalues of the Hamiltonian 
$\hat{H}^{2L}(\varphi)$ 
given by extending 
$H(\varphi)$
to 
$\hat{\Lambda}
:=
\{1, 2, \cdots, 2L\}$
periodically, i.e. 
\[
\hat{t}_{x,x+1} = \cases{
t_{x, x+1}, & $(x=1, \cdots, L)$\cr
t_{x-L, x+1-L}, & $(x = L+1, \cdots, 2L)$\cr},
\quad
\hat{V}=\hat{U}=0.
\]
Once 
(\ref{double})
is proved, 
Theorem \ref{even} leads us to the conclusion
\footnote{(\ref{double}) and Theorem \ref{singlet}
show that the ground state is unique if 
$\varphi= \frac {\pi}{2}, \frac {3 \pi}{2}$ 
which proves the statement in Remark \ref{odd spin}.}.\\
{\it proof of (\ref{double})} : 
By choosing the gauge, we assume 
$\theta_x = 0 (x \ne L), =\varphi (x=L)$.
Let 
$\{ \psi_j^{\varphi} \}_{j=1}^L$ 
be the eigenvector of 
$h(\varphi)$ 
and set 
\[
\hat{\psi}_j^{\varphi}(x) := \cases{
\psi_j^{\varphi}(x), 
&
$(x =1, 2, \cdots, L)$\cr
e^{i \varphi} \psi_j^{\varphi}(x), 
&
$(x = L+1, \cdots, 2L).$\cr}
\]
$\{ \hat{\psi}_j^{\varphi}, \hat{\psi}_j^{\varphi+\pi} \}_{j=1}^L$
are linearly independent and are eigenvectors of 
$\hat{H}^{2L}(2 \varphi)$ 
with eigenvalues 
$\{ e_j (\varphi), e_j (\varphi+\pi) \}_{j=1}^L$. 
Then, 
(\ref{double}) 
follows from the fact that the ground state can be chosen from the 
$S_z = \frac 12$ 
subspace of 
${\cal H}_N$, 
or alternatively, from the theory of one-dimensional periodic Schr\"odinger operators. 
\QED
\begin{remark}
The argument of the above proof shows 
$
\frac {F_n (0) + F_n (\pi)}{2} \ge F_n (\frac {\pi}{2})
$
in general. 
\end{remark}

{\it Proof of Theorem \ref{singlet}}
As usual, 
we work on 
$S_z = 0$ 
subspace. 
We fix 
$\varphi = \left( \frac N2 + 1 \right) \pi$
and write 
$H = H(U)$ 
to specify the 
$U$-dependence of 
$H$. 
For 
$x, y \in {\cal B}_N$, 
let 
$s_{xy} := < x | H(U) | y >$. 
We regard 
$H(U)$ 
as a hopping Hamiltonian on 
${\cal B}_N$ : 
$(H(U) \psi)(x) = \sum_{y \in {\cal B}_N} s_{xy} \psi(y)$. 
Let 
$(H_-(U) \psi)(x) = -\sum_{y \in {\cal B}_N} |s_{xy}| \psi(y)$. 
Then 
by the argument in the proof of Theorem in \cite{N}, 
$H(U)$ 
and 
$H_- (U)$
have same fluxes on each circuit in 
${\cal B}_N$ 
so that they are gauge-equivalent: 
there exists a gauge transformation 
$g$ 
on 
${\cal B}_N$ 
such that 
$H (U) = g^{-1} H_- (U) g$. 
Since 
$s_{xy}$ 
does not depend on 
$U$
for 
$x \ne y$, 
$g$
is independent of 
$U$. 
By Perron-Frobenius theorem,
the ground state 
$\Psi_- (U)$
of 
$H_- (U)$
is unique and so is the ground state 
$\Psi (U)$ 
of 
$H (U)$.
Since
$\Psi_- (U)$ 
and 
$\Psi_- (0)$
are both positive and thus not orthogonal to each other, 
and since 
$\Psi(U)$ 
and 
$\Psi_- (U)$ 
are related via the 
$U$-
independent gauge transformation, 
$\Psi (U)$ 
and 
$\Psi(0)$
have the same spin and thus it suffices to derive the spin of 
$\Psi (0)$. 

Now 
we regard 
$H(0)$
as an operator on 
${\cal H}_N$ 
and let
$e_1 \le e_2 \le \cdots \le e_L$, 
$\psi_1, \psi_2, \cdots, \psi_L
(\in {\bf C}^L)$
be eigenvalues and corresponding eigenvectors of one-particle Hamiltonian of 
$H$ 
(that is, 
$H(0)$ 
as an operator on 
${\cal B}_1$).
Since the ground state 
$\Psi(0)$
of 
$H(0)$
is unique, it is written by
\[
\Psi (0)
=
\prod_{j=1, \sigma = \uparrow, \downarrow}^L
\Psi_{j, \sigma} | \mbox{ vac }>, 
\quad
\mbox{where }
\;
\Psi_{j, \sigma} = \sum_{x = 1}^L 
\psi_{j, \sigma}(x) c_{x, \sigma}^{\dagger}
\]
which has spin zero.
\QED\\

{\it Proof of Theorem \ref{relation}}
(1)
By Theorem \ref{even}(2), 
$E^{\infty}_N (0) = E^{\infty}_N (\pi)$.
The matrix elements of 
$H_{\infty}(\pi)$ 
in terms of the basis 
$B_N$
are non-positive so that Perron-Frobenius theorem shows 
$H_{\infty}(\pi)$ 
has a ground state with 
$S=\frac N2$. 
Hence
$E^{\infty}_N(0) = E^{\infty}_N (\pi) = \sum_{j=1}^N e_j (\pi)$.
Therefore
the statement that 
$H_{\infty}(0)$
does not have a ground state with 
$S= \frac N2$
is equivalent to 
$\sum_{j=1}^N e_j (\pi) < \sum_{j=1}^N e_j (0)$.\\
(2)
The essential ingredient 
of the proof is that the gauge transformation 
$g$, 
which transforms 
$H_{\infty}(\pi)$
to 
$H_{\infty}(0)$, 
transforms the ferromagnetic ground state 
$\Psi_f^{\pi, \infty}$
of 
$H_{\infty}(\pi)$
to those with 
$S< \frac N2$.
In fact, 
$g$
transforms 
$\Psi_f^{\pi, \infty}$
into that which is antisymmetric under the cyclic permutations. 
Let 
${\cal G} :=
\mbox{ Range }P$
be the subspace of 
${\cal H}_N$ 
of states with no doubly occupied sites and let 
${\cal G} = \oplus_{j=1}^K {\cal G}_j $
be the decomposition of 
${\cal G}$
such that 
$H_j (\pi) := H_{\infty} (\pi)|_{{\cal G}_j}$
is irreducible as in the proof of Theorem \ref{even} (2). 
Since the matrix element of 
$H_j (\pi)$
is non-positive, Perron-Frobenius theorem shows that the lowest eigenvector 
$\psi_j (\pi)$ 
is unique and  positive. 
Moreover, 
$H_{\infty} (\pi)$
has a ground state 
$\Psi$
with 
$S=\frac N2$. 
That is, 
there exists 
$\{ a_j \}_{j=1}^K$
such that 
$\Psi = \sum_{j=1}^K a_j \psi_j (\pi)$
is a ground state of 
$H_{\infty} (\pi)$
with  
$S=\frac N2$. 
Fix distinct points 
$x_1, x_2, \cdots, x_N \in \Lambda$.
Let 
\[
A :=
\left\{
c_{y_1, \sigma_1}^{\dagger}
\cdots
c_{y_N, \sigma_N}^{\dagger} | \mbox{vac}> \in {\cal G} : 
y_i = x_1, \cdots, x_N, \sigma = \uparrow, \downarrow \} 
\right\}, 
\;
A_j := A \cap  B_j
\]
and let 
$P_A$
be the orthogonal projection onto the subspace of 
${\cal G}$
spanned by 
$A$. 
Since 
$\Psi$
has 
$S=\frac N2$, 
$P_A \Psi = a \sum_k \rho_k$, 
$\rho_k \in A$
for some 
$a > 0$ 
which implies 
$P_A \psi_j(\pi) = b_j \sum_k \nu_{jk}$, 
$\nu_{jk} \in A_j$
for some 
$b_j > 0$. 
We normalize 
$\psi_j (\pi)$
such that
$b_j = 1$.
Since
$H_{\infty} (\pi)$
and 
$H_{\infty} (0)$
are gauge equivalent, 
there exists a gauge transformation 
$g$
such that 
$\psi_j (0) = g \psi_j (\pi)$. 
Suppose 
$H_{\infty} (0)$
has a ground state
$\tilde{\Psi}$
with 
$S=\frac N2$. 
Then 
$\tilde{\Psi} = \sum_{j=1}^K c_j \psi_j(0)$
for some 
$\{ c_j \}_{j=1}^K$ 
and 
\begin{equation}
{\bf S}^2 ( P_A \tilde{\Psi} )
=
\frac N2 
\left(\frac N2+1\right) ( P_A \tilde{\Psi}).
\label{ferro}
\end{equation}
Let 
$H_{\infty} ^0 (\varphi)$
be the Hamiltonian with 
$| t_{x,x+1} | =1$
and 
$V=0$. 
Let 
$\psi_j^0 (\pi)$
be the corresponding lowest eigenvector of 
$H_{\infty}^0 (\pi)|_{{\cal G}_j}$. 
Normalize 
$\psi_j^0 (\pi)$
by the same procedure as above.
Since 
$H_{\infty}^0 (\pi)$
is transformed to 
$H_{\infty}^0 (0)$
by the same gauge transformation 
$g$,
\begin{equation}
P_A \psi_j^0 (0) = P_A \psi_j (0).
\label{same}
\end{equation}
On the other hand, 
$\tilde{\Psi}_0 :=
\sum_{j=1}^K c_j \psi_j^0 (0)$
is a ground state of 
$H(0)$
which satisfies 
\begin{equation}
{\bf S}^2 (P_A \tilde{\Psi}_0 )
=
\frac N2 \left(\frac N2 + 1 \right)
(P_A \tilde{\Psi}_0)
\label{contradiction}
\end{equation}
by 
(\ref{ferro}), (\ref{same}).
(\ref{contradiction})
contradicts to the fact that 
$H_{\infty}^0 (0)$
has no ground state with 
$S = \frac N2$,
since we have 
$\sum_{j=1}^N e_j (\pi)<\sum_{j=1}^N e_j (0)$
in this case.
\QED\\

{\it Proof of Theorem \ref{connection}}
Let 
$U<\infty$. 
Then 
there is a gauge transformation 
$g$
which is independent of 
$U$
such that 
$H(0) = g H_{PF} g^{-1}$. 
$H_{PF}$
is the Hamiltonian whose matrix elements 
(in terms of 
$B_N$)
are non-positive and have the same absolute values as those of 
$H(0)$.
The ground states 
$\Psi_s^{0}, \Psi_{PF}$
of 
$H(0), H_{PF}$
satisfy 
\begin{equation}
\Psi_s^{0} = g \Psi_{PF}
\label{equation}
\end{equation}
and 
$\Psi_s^{0}$
has 
$S=0$
while 
$\Psi_{PF}$
is positive.
When 
$U$
goes to infinity, 
$\lim_{U \uparrow \infty} \Psi_s^{0} = \Psi_s^{0, \infty}$
which is a singlet ground state of 
$H_{\infty}(0)$.
On the other hand, 
$\lim_{U \uparrow \infty} H_{PF} = H_{\infty}(\pi)$
and moreover, 
$\lim_{U \uparrow \infty} \Psi_{PF} = \Psi_f^{\pi, \infty}$
where 
$\Psi_f^{\pi, \infty}$
is the ferromagnetic ground state of 
$H_{\infty}(\pi)$. 
This follows 
from the observation that both 
$\Psi_{PF}$ 
and 
$\Psi_f^{\pi, \infty}$
are positive and the other ground states of 
$H_{\infty}(\pi)$
are not non-negative. 
Letting 
$U \to \infty$
in 
(\ref{equation}), 
we have 
\[
\Psi_s^{0, \infty} = g_{\infty} \Psi_f^{\pi, \infty},
\quad
g_{\infty} = g|_{\cal G}
\]
which is the desired conclusion.
\QED
%
\section{Discussion}
In this paper, 
we study the flux phase problem, that is, to minimize the ground state energy w.r.t. the flux, in the one-dimensional  many particle systems. 
In particular, 
we study the case in which the particle number is odd at half-filling, and deduced that the optimal flux is 
$\frac {\pi}{2}, \frac {3 \pi}{2}$, 
in the absence of on site interaction.
Such results are already derived by the Bethe ansatz calculation \cite{YF}, and thus our contribution is to show that this is also true even if the hopping coefficients are not constant, namely the hopping disorder has no effect on the optimal flux.
Moreover, 
unlike the case of even number of particles, 
we find something unusual happens: 
Theorem \ref{odd} is not necessarily true if 
$V \ne 0$, 
implying that the potential disorder may have some effect on the optimal flux, 
or if 
the temperature is nonzero 
(Remarks \ref{potential disorder},\ref{nonzero temperature}).
This also implies 
the method of proof of Theorem in \cite{N} may not apply to the case of odd number of particles in general. 

Next, 
we study the spin of the ground state and showed that it is zero when the flux is optimal. 
When it is not optimal, 
the spin is not zero and changes its value depending on the hopping coefficients $T$, the on-site interaction 
$U$, and the external potential
$V$, implying it is not stable. 
It also implies 
the conclusion of Lieb-Mattis theorem is not true for such cases so that the boundary effect is not negligible.
Nevertheless, 
if the flux is optimal, the spin is always zero for any 
$T$, $U$,
and 
$V$, 
implying that it is always stable under the perturbation.

Moreover, 
we study the case in which 
$U = \infty$
and found a relation between the spin of the ground state and the sum of the lowest eigenvalues of the one-particle Hamiltonian. 
Since 
the spin is a `robust' property, we can derive some information on the sum of lowest eigenvalues which holds for any $T$, $U$, and $V$. 
We also discussed the `spiral state' : 
a singlet ground state of 
$H_{\infty}(0)$ 
which is obtained by a simple gauge transformation of a ferromagnetic state of $H_{\infty}(\pi)$.
These results 
seem to reveal interesting connection between the flux threading the system and spin of the ground state. 
%

\section{Appendix : proof of Lemma 2.1}
Let 
$\Psi_1$
be the eigenvector of 
$H_j^p$
with eigenvalue 
$E$. 
It suffices to construct the eigenvector  
$\Psi_0$
of 
$H_k^N$
with the same eigenvalue 
$E$.
We write  
$\Psi_1, \Psi_0$
in terms of the linear combination of their basis: 
\begin{eqnarray*}
\Psi_1
&=&
\sum_{
x_1, \cdots, x_N,
\sigma_1, \cdots, \sigma_N
}
a(x_1, \sigma_1; x_2, \sigma_2; \cdots ; x_N, \sigma_N)
c_{x_1, \sigma_1}^{\dagger}
c_{x_2, \sigma_2}^{\dagger}
\cdots
c_{x_N, \sigma_N}^{\dagger}
| \mbox{vac}>
\\
\Psi_0
&=&
\sum_{x_1, \cdots, x_N, \sigma_1, \cdots, \sigma_N}
b(x_1, \sigma_1; x_2, \sigma_2; \cdots ; x_N, \sigma_N)
c_{x_1, \sigma_1}^{\dagger}
c_{x_2, \sigma_2}^{\dagger}
\cdots
c_{x_N, \sigma_N}^{\dagger}
| \mbox{vac}>
\end{eqnarray*}
where in 
$\Psi_1$, 
$c_{x_1, \sigma_1}^{\dagger}
c_{x_2, \sigma_2}^{\dagger}
\cdots
c_{x_N, \sigma_N}^{\dagger}
| \mbox{ vac}> \in B_j^p$
and similarly for 
$\Psi_0$. 
Fix 
$x_1 < x_2 < \cdots < x_N$. 
Pick any spin configuration 
$(\sigma_1, \sigma_2, \cdots, \sigma_N)$
and 
we determine
$b(x_1, \sigma_1; x_2, \sigma_2; \cdots ; x_N, \sigma_N)
$
by the following steps. 
Pick any fixed element 
$c_{x_1, \tau_1}^{\dagger}
c_{x_2, \tau_2}^{\dagger}
\cdots
c_{x_N, \tau_N}^{\dagger} | \mbox{vac}>
\in B_k^N$.
Then for any other elements 
$c_{x_1, \sigma_1}^{\dagger}
c_{x_2, \sigma_2}^{\dagger}
\cdots
c_{x_N, \sigma_N}^{\dagger} | \mbox{vac}>
\in B_k^N$, 
$(\sigma_1, \sigma_2, \cdots, \sigma_N)$
is the cyclic permutation of 
$(\tau_1, \tau_2, \cdots, \tau_N)$
and since 
$B_k^N$
has period 
$N$, 
we can find 
$k$
($1 \le k \le N$)
uniquely such that 
$(\sigma_1, \sigma_2, \cdots, \sigma_N)
=
(\tau_k, \tau_{k+1}, \cdots, \tau_N, \tau_1, \tau_2, \cdots, \tau_{k-1})$. 
Pick and fix any element
$c_{x_1, \tau'_1}^{\dagger}
c_{x_2, \tau'_2}^{\dagger}
\cdots
c_{x_N, \tau'_N}^{\dagger} | \mbox{vac}>
\in B_j^p$.
We define 
$b(x_1, \sigma_1; x_2, \sigma_2; \cdots ; x_N, \sigma_N)
$
as 
\[
b(x_1, \sigma_1; x_2, \sigma_2; \cdots ; x_N, \sigma_N)
:=
a(x_1, \tau'_k ; x_2, \tau'_{k+1}; \cdots ; x_N, \tau'_{k-1}).
\]
It is straightforward to check 
$\Psi_0$
is the eigenvector of 
$H_j^N$
with eigenvalue 
$E$. 
Lemma \ref{boson} is proved.

%


\small
\begin{thebibliography}{99}
\bibitem{K}
Kusmartsev, F. V.,
 Magnetic resonance 
on a ring of aromatic molecules, 
J. Phys. Condensed Matter {\bf 3}(1991), 3199-3204. 
%
\bibitem{L1}
Lieb, E. H., 
Two Theorems on the Hubbard model,
Phys. Rev. Lett. {\bf 62}(1989), 1201-1204. 
%
\bibitem{L2}
Lieb, E. H., 
Flux Phase of the Half-Filled Band,
Phys. Rev. Lett. {\bf 73}(1994), 2158-2161. 
%
\bibitem{LL}
Lieb, E. H., and Loss, M., 
 Fluxes, Laplacians and Kasteleyn's theorem,
Duke Math. J. {\bf 71}(1993), 337-363. 
%
\bibitem{LM}
Lieb, E. H., and Mattis, D. C., 
 Theory of ferromagnetism 
and the ordering of electronic levels,
Phys. Rev. {\bf 125}(1962), 164-172. 
%
\bibitem{LN}
Lieb, E. H., and Nachtergaele, B., 
 Stability of the Peierls instability 
for ring-shaped molecules,
Phys. Rev. {\bf B51}(1995), 4777-4791. 
%
\bibitem{N}
Nakano, F.: 
The flux phase problem on the ring, 
 J. Phys. {\bf A 33} (2000), no. 30, 5429--5433. 
%
\bibitem{NN}
Nakano, F., and Nomura, Y.:
Random magnetic fields on line graphs, 
J. Math. Phys. {\bf 44}(2003), pp. 4988-5002.
%
\bibitem{R}
Rokhsar, D.:
Solitons in Chiral-Spin Liquids, 
Phys. Rev. Lett. {\bf 65}(1990), 1506-1509.
%
\bibitem{YF}
Yu, F., and Fowler, M., 
 Persistent current of a Hubbard ring 
threaded with a magnetic flux,
Phys. Rev. {\bf B45}(1992), 11795-11804.
%
\end{thebibliography}
\end{document}